\documentclass[preprint2]{aastex}

\def\be{\begin{equation}}
\def\ee{\end{equation}}

\begin{document}

\title{Aristotelian Electrodynamics solves the Pulsar: Lower Efficiency of Strong Pulsars} 

\author{Andrei Gruzinov}

\affil{CCPP, Physics Department, New York University, 4 Washington Place, New York, NY 10003}

\begin{abstract}

In Aristotelian Electrodynamics (AE), due to radiation over-damping, the velocity rather than the acceleration of a charge is determined by the local electromagnetic field. Treating electrons and positrons separately, AE seems to give a faithful description of the flow of charges in a pulsar magnetosphere. AE might allow to calculate the pulsar emission in full detail, at all energies except the radio.

But AE is not a full theory -- the plasma production processes must be added. Here we show that the pulsar magnetosphere and the resulting emission are sensitive to the plasma production rate near the light cylinder. In strong pulsars (high production rate, due to photon-photon collisions), the number of field lines closing beyond the light cylinder decreases. This reduces both the pulsar efficiency (the ratio of the pulsed bolometric luminosity to the spin-down power) and the characteristic photon energy, in overall agreement with the Fermi data.

For weak pulsars (low production rate), our numerical implementation of AE works stably and seems to be ready to calculate the lightcurves and spectra.  But for strong pulsars, the code (included) needs improvement or better understanding.

~~~

~~~

\end{abstract}

\section{Introduction}

To ``solve the pulsar'', one needs to describe the production, motion, and radiation of electrons and positrons. Aristotelian Electrodynamics (AE) describes the motion and curvature radiation of charges in a strong electromagnetic field. AE is applicable to pulsars. 

AE appears to be ready (after going 3D, high resolution) to calculate the lightcurves and spectra of weak pulsars (insignificant pair production near the light cylinder). For strong pulsars (high rates of pair production near the light cylinder), one needs to add the relevant QED kinetics. This seems to be doable (\S4.3), but there is another problem. At high production rates, our AE code behaves in an unclear manner. Since the code was written by a numerical amateur, there must be room for improvement. 

In \S2 we describe Aristotelian Electrodynamics, in \S3 we use AE to calculate the pulsar magnetosphere, in \S4 we discuss how AE can be used to calculate the pulsar emission.

\section{Aristotelian Electrodynamics} 

In pulsar-strength electromagnetic field, because of strong radiation damping, velocities rather than accelerations of positrons and electrons are given by the local electromagnetic field: 
\be\label{ae}
{\bf v}_{\pm}={{\bf E}\times {\bf B}\pm(B_0{\bf B}+E_0{\bf E})\over B^2+E_0^2}.
\ee
Here the scalar $E_0$ and the pseudoscalar $B_0$ are the proper electric and magnetic fields defined by 
\be
B_0^2-E_0^2=B^2-E^2,~ B_0E_0={\bf B}\cdot {\bf E},~ E_0\geq 0.
\ee
Equation (\ref{ae}) gives unit velocities (speed of light $c=1$). The charges actually move slightly slower than light (see \S4). \footnote{More precisely: (i) where $E_0$ is large, the charges move at high Lorentz factors, (ii) where $E_0$ is small, the sign of $B_0$ frequently changes and the charges move at an average velocity below $c$.}

Equation (\ref{ae}) says that in the frames where ${\bf E}$ is parallel to ${\bf B}$, the charges move at the speed of light parallel and antiparallel to ${\bf E}$. As they move along (\ref{ae}), the charges are pulled by $E_0$ and dragged by the curvature radiation. We will assume that a charge reaches the terminal Lorentz factor before it moves over the characteristic length scale of the field. For pulsars, where all the action is near the light cylinder (see below), this requirement of instantaneous parallel dynamics reads
\be\label{app}
L_{sd}\gg L_e\left( {R\over r_e} \right) ^{2/3},
\ee
where $L_{sd}$ is the spin-down power, $R={c\over \Omega}$ is the light cylinder radius, $\Omega$ is the angular velocity of the star, $r_e={e^2\over mc^2}=2.8\times 10^{-13}$cm is the classical electron radius, $L_e={mc^2\over r_e/c}=8.7\times 10^{16}$erg/s is the ``classical electron luminosity''. In astrophysical notation, the applicability condition (\ref{app}) reads
\be\label{appa}
L_{34}\gg 5.7\times 10^{-5} P_{ms}^{2/3},
\ee
where $L_{34}$ is the spin-down power in units of $10^{34}$erg/s and $P_{ms}$ is the pulsar period in ms. The applicability condition (\ref{appa}) is satisfied, and by a large margin, by all pulsars in the Fermi catalog (Abdo et al 2010).

\section{Using AE to describe pulsars} 

\subsection{AE equations}

To calculate the pulsar magnetosphere, one needs to solve Maxwell equations. Maxwell equations can be solved (numerically) if one knows the electric current ${\bf j}$. According to AE, the current is 
\be\label{cur}
{\bf j}=\rho_+{\bf v}_+-\rho_-{\bf v}_-,
\ee
where $\rho_+$ is the positron charge density, $\rho_-$ is the absolute value of the electron charge density, and ${\bf v}_{\pm}$ is given by (\ref{ae}).

The charge densities should be calculated from the continuity equations
\be\label{con}
\dot{\rho_{\pm}}+\nabla \cdot (\rho_{\pm}{\bf v}_{\pm})=Q,
\ee
where $Q$ is the plasma production rate.

\subsection{Previous AE calculation of pulsars}

In our previous AE calculation of the pulsar (Gruzinov 2012a), instead of modeling the plasma production $Q$, we postulated an expression for the charge-normalized number density $\rho _0$, 
\be
\rho _0\equiv \rho_++\rho_-.
\ee
(We used $\rho _0=\sqrt{\rho ^2+f^2(B^2+E^2)/r^2}$, where $\rho =\rho_+-\rho_-$ is the electric charge density, $r$ is the spherical radius, and $f$ is a dimensionless parameter which we varied.)

Once the plasma density $\rho _0$ is postulated, equation (\ref{cur}) becomes an Ohm's law -- an expression giving ${\bf j}$ in terms of the electromagnetic field --
\be\label{ohm}
{\bf j}={\rho {\bf E}\times {\bf B}+\rho _0(B_0{\bf B}+E_0{\bf E})\over B^2+E_0^2},
\ee
\be
\rho =\nabla \cdot {\bf E}.
\ee
Now the magnetosphere can be calculated in terms of the electromagnetic degrees of freedom only.

The result of the calculation was non-trivial. It turned out, that the magnetosphere hardly depends on the assumed ``fiducial multiplicity'' $f$ . In particular, we claimed to have calculated the universal value of the pulsar efficiency (the ratio of the pulsed bolometric luminosity to the spin-down power):
\be
\epsilon \approx {0.5\over 1+5\sin ^2\theta },
\ee
where $\theta$ is the spin-dipole angle. The resulting $\theta$-averaged efficiency $\epsilon\sim 0.1$ is in an overall agreement with the data, especially if one ignores brighter pulsars. Since in AE the drag comes from the curvature radiation, one can also give a crude estimate of the photon energy (see \S4) $E\sim c^{5\over 8}\hbar e^{-{3\over 4}}L_{sd}^{3\over 8}R ^{-{1\over 4}}\sim 3L_{34}^{3\over 8}P_{ms}^{-{1\over 4}}{\rm GeV}$. This again roughly agrees with the data if the brighter pulsars are ignored.

\subsection{Why the postulated multiplicity model is wrong}

Given the insensitivity to the assumed multiplicity, one might think that the actual plasma production rate is irrelevant. But it is easy to see that this cannot be right. 

Suppose the plasma production rate near the light cylinder $Q$ is high,
\be
Q\gg \Omega \rho _{GJ},
\ee
where $\rho _{GJ}\sim B/R$ is the Goldreich-Julian (1969) density, $B$ is the magnetic field at the light cylinder. What is the fate of all these constantly added charges?

\begin{figure}
\plotone{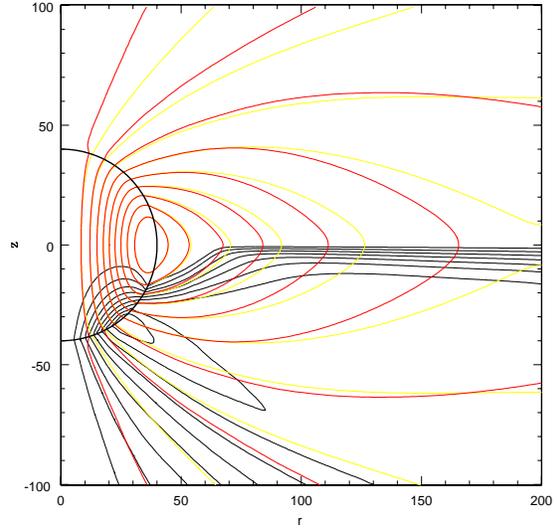}
\caption{Weak pulsar -- no plasma production near the light cylinder. The star rotates at half the speed of light. Thin black -- poloidal current isolines. Thicker yellow -- poloidal magnetic field. Thick red -- electric potential.}
\end{figure}

\begin{figure}
\plotone{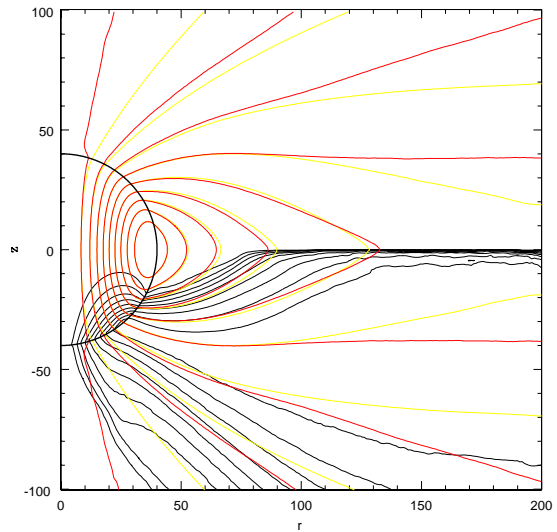}
\caption{Stronger pulsar -- significant plasma production near the light cylinder. }
\end{figure}

The charges cannot be removed faster than light, and the plasma density $\rho _0$ will be much higher than $\rho _{GJ}$. The proper electric field $E_0$ has to be small in most of the volume -- otherwise the damping rate $q={\bf j}\cdot {\bf E}=c\rho _0E_0$ would be much greater than $L_{sd}/R^3$. At small $E_0$, and considering the axisymmetric case for simplicity, we see from (\ref{ae}) that the poloidal motion of the charges is along the poloidal magnetic field. Then at least one of the species has to move into the equatorial current layer (see Fig.1). Once inside the current layer, this species will have to experience the full potential drop of the current layer. This again results in more damping power than $L_{sd}$.

The above argument is not in contradiction with the model of (Gruzinov 2012b), where we show that certain plasma production rates $Q$ do give a fixed-multiplicity AE.  These models require pair annihilation -- negative $Q$ values must be allowed. Then both species move into the equatorial current layer and get annihilated there. But in real pulsars the annihilation rate should be very small, and such models should not be relevant.

\subsection{New AE calculation of pulsars}

For the aligned rotator, we have numerically integrated the full axisymmetric AE system: (\ref{ae}, \ref{cur}, \ref{con}) plus Maxwell equations
\be\label{mab} 
\dot{\bf B}=-\nabla \times {\bf E},
\ee
\be\label{mae}
\dot{\bf E}=\nabla \times {\bf B}-{\bf j}.
\ee
The plasma production was
\be
Q=\alpha _1Q_1+\alpha _2Q_2,
\ee
where $Q_1$ models the avalanche plasma production near the star (Ruderman and Sutherland 1975) and $Q_2$ models the plasma production by photon-photon collisions near the light cylinder.

For no particular reason we used
\be
Q_1=\theta (1.4r_s-r)\theta (E_0-E_c) (1-{E_c^2\over E_0^2}),
\ee
\be\label{q2}
Q_2={(\Omega r)^2\over 1+(\Omega r)^6},
\ee
where $r$ is the spherical radius, $r_s$ is the radius of the star, $E_c$ is the critical field for the near-star avalanche, $\Omega$ is the angular velocity of the star. Further details of the model are given in the Appendix. 

With no plasma production near the light cylinder (Fig.1), we get a magnetosphere similar to (Gruzinov 2012b). Since the equatorial current experiences a large potential drop, this pulsar will have a high efficiency. 

For significant plasma production near the light cylinder (Fig.2), the number of field lines closing beyond the light cylinder decreases. The equatorial current experiences a smaller potential drop, this pulsar will have a lower efficiency. The calculated damping rate, $\int d^3r~(\rho_++\rho_-)E_0$, indeed drops by about a factor of 2. 

At low plasma production, the simulation results seem to converge just as nicely as in a purely electromagnetic simulation of (Gruzinov 2012ab). But at significant plasma production, our code probably needs improvement or better understanding. The convergence of the code (with time step and resolution) is unclear to the author. The results of the calculation given in Fig.2 have the status of an illustration. But since the underlying physics seems to be right, we decided to show this illustration. 

How small can the pulsar efficiency become at high multiplicity? A crude estimate of \S4 suggests that efficiency $\epsilon$ can't decrease too much below the ideal value of $\epsilon \sim 0.1$.  Observationally, even the strongest pulsar (Crab) has efficiency $\epsilon \sim 0.01$ -- just a factor of 10 below the mean ideal value (X-ray dominated; the light cylinder of Crab can be opaque to high-energy gamma-rays). 

From a purely theoretical perspective, the minimal efficiency problem is unclear to the author. Our results seem to indicate that at high plasma production rates the AE magnetosphere approaches the standard zero-efficiency force-free magnetosphere of Contopoulos et al (1999) and Spitkovsky (2006). But one needs to remember that the ``periodic axisymmetric pulsar'' (Gruzinov 2011) has a 100\% efficiency even at an arbitrarily high multiplicity.

\section{Pulsar emission}

According to AE, the Poynting energy is extracted at the current layer just beyond the light cylinder. AE predicts the terminal Lorentz factor of the charges $\gamma$:
\be
ceE_0={2\over 3}{e^2c\over R_c}\gamma ^4,
\ee 
where $R_c$ is the radius of curvature of the trajectory of a charge. $R_c$ is also predicted by AE. Knowing $\gamma$ and $R_c$, one calculates the spectrum of the primary gamma-rays (usually $\sim$ GeV ). 

In weak pulsars, the primary gamma-rays must be dominating the emission, while being irrelevant for the plasma production -- the plasma is produced only by the near-star avalanche. In strong pulsars, the situation seems to be more complicated.

\subsection{Weak pulsars}

The Fermi catalog lists 6 pulsars with $P_{ms}\sim 300$ and $L_{34}\sim 1$. These have the photon cutoff energy $\sim$Gev. For 3 of these pulsars, the data of Possenti et al (2002) gives a crude upper bound for the X-ray luminosity. Neglecting the geometric factors, one then gets a crude estimate for the plasma production near the light cylinder. Comparing to the Goldreich-Julian density, we conclude that these pulsars are weak.

These pulsars still have copious pair production near the star. Since the plasma properties near the light cylinder seem to be insensitive to how exactly the near-star avalanche occurs, AE can be used directly to predict the lightcurves and spectra. It is already known (Bai, Spitkovsky 2010) that postulating gamma-ray emission from the current layer beyond the light cylinder gives the lightcurves crudely consistent with observations. 

\subsection{Millisecond pulsars}

Estimating pair production near the light cylinder for millisecond pulsars in the same manner as above, we also classify them as weak. But the near-star avalanche in millisecond pulsars probably requires careful modeling, because by the usual criteria (Ruderman, Sutherland 1975) the avalanche is barely going. The near-star avalanche can noticeably contribute to the observed emission (although the current layer emission cannot be subdominant).

\subsection{Strong Pulsars}

Below we describe one scenario for what might be happening near the light cylinder in strong pulsars -- the ``light-cylinder avalanche''. In this scenario, careful modeling of kinetics seems to be a must for predicting the emission.

For the sake of the argument, take the light cylinder radius $R\sim 3\times 10^8$cm and the magnetic field near the light cylinder $B\sim100$kG. This would be a pulsar with $L_{sd}\sim 3\times 10^{37}$erg/s. Let $\epsilon _1L_{sd}$ be the primary gamma-ray luminosity at about 1GeV. Neglecting geometry, we estimate the 1GeV gamma-ray density at the light cylinder 
\be
n_\gamma \sim {\epsilon _1L_{sd}\over 4\pi R^2c(1{\rm GeV})}\sim 10^{12}\epsilon _1{\rm cm}^{-3}.
\ee

Now send a test 1keV X-ray photon into the light cylinder region. With a probability
\be
p\sim Rr_e^2n_\gamma \sim 3\times 10^{-5}\epsilon _1
\ee
the test X-ray photon will produce a pair -- an electron and a positron of $\sim$GeV energy. In the 100kG magnetic field, the pair will emit $N\sim 10^6$ X-ray photons of $\sim$keV energy . If $pN>1$, we get an avalanche.

In the simplest model, the pulsar balances at the threshold of the light-cylinder avalanche, $\epsilon \sim \epsilon _c\sim 0.03$. For $\epsilon > \epsilon _c$, copious pair production near the light cylinder opens the field lines (Fig.2.) and the efficiency $\epsilon$ drops. For $\epsilon < \epsilon _c$, the avalanche disappears, pairs are no longer produced near the light cylinder and the field tries to return to the high-efficiency weak-pulsar configuration (Fig.1).

\appendix

\section{Numerical AE in axial symmetry}

The code used to generate the figures can be found at http://cosmo.nyu.edu/andrei/AE/. 

We numerically integrate the AE equations (\ref{ae}, \ref{cur}, \ref{con}, \ref{mab}-\ref{q2}) with diffusive regularizations. It would be good to have a code which gets rid of the artificial diffusivities $\beta$ and $D$ (see below).

\subsection{Variables and Fields}

We assume axial symmetry. The coordinates are 

\begin{itemize}

\item $t$ -- time

\item $(r,\theta ,z)$ -- cylindrical coordinates

\item $R=\sqrt{r^2+z^2}$ -- spherical radius

\end{itemize}

The fields are

\begin{itemize}

\item $\rho _\pm$ -- positron/electron density (charge normalized), defined for $R>r_s$

\item ${\bf j}_\pm$ -- positron/electron current, defined for $R>r_s$

\item ${\bf j}$ -- electric current

\item ${\bf E}$, ${\bf B}$ -- EM field

\end{itemize}

\subsection{Parameters}
The parameters of the model are
\begin{itemize}

\item $r_s$ -- the radius of the star

\item $\Omega$ -- the angular velocity of the star

\item $\sigma _s$ -- the electrical conductivity of the star

\item ${\bf j}_e$ -- permanent external current in the star (in the code, we use a purely toroidal current which scales linearly with $r$)

\item $\alpha _1$, $\alpha _2$ -- the pair production rates near the star and at the light cylinder

\item $E_c$ -- the critical electric field for the inner pair production 

\item $\beta$ -- ``toroidal diffusivity'' of the EM field

\item $D$ -- diffusivity of charges (in the code, we use diffusivity which scales linearly with $R$)

\end{itemize}

\subsection{Equations}

The AE equations read

\be
\begin{array}{l}
\partial _t\rho _\pm +r^{-1}\partial _r(rj_{\pm r})+\partial _z(j_{\pm z})=Q , ~~~~~~~~~~~~~~ R>r_s,\\
\\
\partial _tB_r=\partial _zE_\theta ,\\
\partial _tB_\theta=-\partial _zE_r+\partial _rE_z+\beta \left(r^{-1}\partial _r(r\partial _rB_\theta )+\partial _z^2B_\theta -r^{-2}B_\theta \right) ,\\
\partial _tB_z=-r^{-1}\partial _r(rE_\theta ) ,\\
\\
\partial _tE_r=-\partial _zB_\theta-j_r ,\\
\partial _tE_\theta=\partial _zB_r-\partial _rB_z-j_\theta+\beta \left(r^{-1}\partial _r(r\partial _rE_\theta )+\partial _z^2E_\theta -r^{-2}E_\theta \right) ,\\
\partial _tE_z=r^{-1}\partial _r(rB_\theta )-j_z ,\\
\\
{\bf j}={\bf j}_e+\sigma _s\left( {\bf E}+\Omega r\hat{\theta}\times {\bf B} \right), ~~~~~~~R<r_s ,\\
{\bf j}={\bf j}_+-{\bf j}_- ,~~~~~~~~~~~~~~~~~~~~~~~~~~~~~~R>r_s, \\
{\bf j}_\pm=\rho _\pm{\bf v}_\pm -D\nabla \rho _\pm ,\\
\\
{\bf v}_{\pm}={{\bf E}\times {\bf B}\pm(B_0{\bf B}+E_0{\bf E})\over B^2+E_0^2+0} ,~~~B_0^2-E_0^2=B^2-E^2,~ B_0E_0={\bf B}\cdot {\bf E},~ E_0\geq 0,\\
\\
Q=\alpha _1Q_1+\alpha _2Q_2 ,\\
Q_1=\theta (1.4r_s-R)\theta (E_0-E_c) (1-{E_c^2\over E_0^2}),\\
Q_2={(\Omega R)^2\over 1+(\Omega R)^6}.

\end{array}
\ee

Note: at high $\sigma _s$, one can use a non-relativistic Ohm's law inside the star. For us the star is just a black box of high conductivity, and the detailed form of conductivity is irrelevant.

\subsection{Initial and Boundary Conditions}

At $t=0$ all fields are equal to zero. 

We solve the problem inside a cylinder $r<a$, $|z|<b$. The boundary conditions are absorbing for the density and outgoing for the EM field

\be
\begin{array}{l}
\rho _\pm=0, ~~~R=r_s, \\
\rho _\pm=0, ~~~r=a,\\
\rho _\pm=0, ~~~|z|=b,\\
\\
E_r=-B_\theta, ~~~z=-b,\\
E_\theta =B_r, ~~~z=-b,\\
\\
E_r=B_\theta, ~~~z=b,\\
E_\theta =-B_r, ~~~z=b,\\
\\
E_\theta =B_z, ~~~r=a,\\
E_z =-B_\theta, ~~~r=a,\\
\end{array}
\ee

Note: because of the diffusive term in the $B_\theta$ equation, (A1), (A2) is not a correctly posed problem. But since $\beta$ is just a regularization, we coded (A1), (A2) in a way which required no (explicit) additional boundary conditions. 

\end{document}